\documentclass[conference]{IEEEtran}

\ifCLASSINFOpdf
\else
\fi
\usepackage{multicol}
\usepackage{tikz}
\usepackage{tkz-graph}
\usepackage{bbm}
\usepackage{booktabs}
\usepackage{algpseudocode}
\usepackage{algorithm}

\usepackage{color}
\usepackage[cmex10]{amsmath}
\usepackage{dblfloatfix}
\usepackage{amsfonts}
\usepackage{url}
\usepackage{pgfplots}
\usepackage{pgfplotstable}
\usepackage{hyperref}
\hypersetup{
    colorlinks=true,
    linkcolor=blue,
    filecolor=magenta,      
    urlcolor=cyan,
}
\usepackage{listings}

\begin{document}
\pagenumbering{gobble}

%
\title{\textbf{\Large A Scalable Database for the Storage of Object-Centric Event Logs (Extended Abstract)\\[-1.5ex]}}

\author{
\IEEEauthorblockN{~\\[-0.4ex]\large Alessandro Berti\IEEEauthorrefmark{1}, Anahita Farhang Ghahfarokhi\IEEEauthorrefmark{1}, Gyunam Park\IEEEauthorrefmark{1}, Wil M.P. van der Aalst\IEEEauthorrefmark{1}}
\IEEEauthorblockA{\IEEEauthorrefmark{1}Process and Data Science Department, RWTH Aachen University\\
Process and Data Science department, Lehrstuhl fur Informatik 9 52074 Aachen, Germany\\
Emails: {a.berti@pads.rwth-aachen.de}, {wvdaalst@pads.rwth-aachen.de}}
}

\maketitle

\begin{abstract}
Object-centric process mining provides a set of techniques for the analysis of event data where events are associated to several objects.
To store Object-Centric Event Logs (OCELs), the JSON-OCEL and JSON-XML formats have been recently proposed. However, the proposed implementations of the OCEL are file-based. This means that the entire file needs to be parsed in order to apply process mining techniques, such as the discovery
of object-centric process models. In this paper, we propose a database storage for the OCEL format using the MongoDB document database.
Since documents in MongoDB are equivalent to JSON objects, the current JSON implementation of the standard could be translated straightforwardly
in a series of MongoDB collections.
\end{abstract}




%
\IEEEpeerreviewmaketitle

\pgfplotstableread[col sep=space,row sep=newline,header=true]{
x   y
6.0 1.38
7.0 2.46
8.0 3.54
}\scalabilityDFGcalc

\pgfplotstableread[col sep=space,row sep=newline,header=true]{
x   y
6.0 2.77
7.0 3.60
8.0 3.81
}\memoryOccupation

\section{Significance of the Tool}

OCEL \url{http://www.ocel-standard.org/1.0/specification.pdf} has been proposed to model the structure of object-centric event logs \cite{ghahfarokhi2021ocel}.
Implementations of the format have been made available for JSON and XML file formats, and tool support is proposed for the Python and Java languages.
For all these, the event log is stored in a JSON/XML file that can be ingested in-memory by the tools/libraries. The necessity to load the log in-memory
makes it difficult to manage a huge amount of object-centric event data since memory is a limited asset.
With this paper, a novel implementation of the format is proposed based on the MongoDB document database. Documents can be imported in MongoDB starting from JSON objects.
Hence, the JSON-OCEL implementation could be translated easily to MongoDB. Moreover, MongoDB can mix in-memory and on-disk computations to provide efficient data science pipelines.
Other advantages of MongoDB that we exploit are:
the fine-grained support for indexes (i.e., multikey), which makes ad-hoc querying faster; the fine-grained support for aggregations (i.e., grouping) that permits
to move some of the computations at the database level; the support to replication, which provides redundancy and increases data availability \url{https://docs.mongodb.com/manual/replication/}.
Graph databases have been assessed previously for the storage of object-centric event data \cite{esser2021multi,jalali2020graph,esser2019storing}, but the direct translation of the specification of OCEL in a graph database is more challenging\footnote{Even if object-centric event logs can, in general, be uploaded to graph databases as shown in \url{https://doi.org/10.5281/zenodo.3865221}}. Also, columnar storages have been used \cite{wang2020cloud,berti2018extracting}, with the limitations
that they work for basic column types but do not provide comprehensive support to JSON and advanced data types.

\section{Main Features of the Tool}

\begin{figure}[ht]
\vspace{-0.5cm}
\includegraphics[width=0.92\columnwidth]{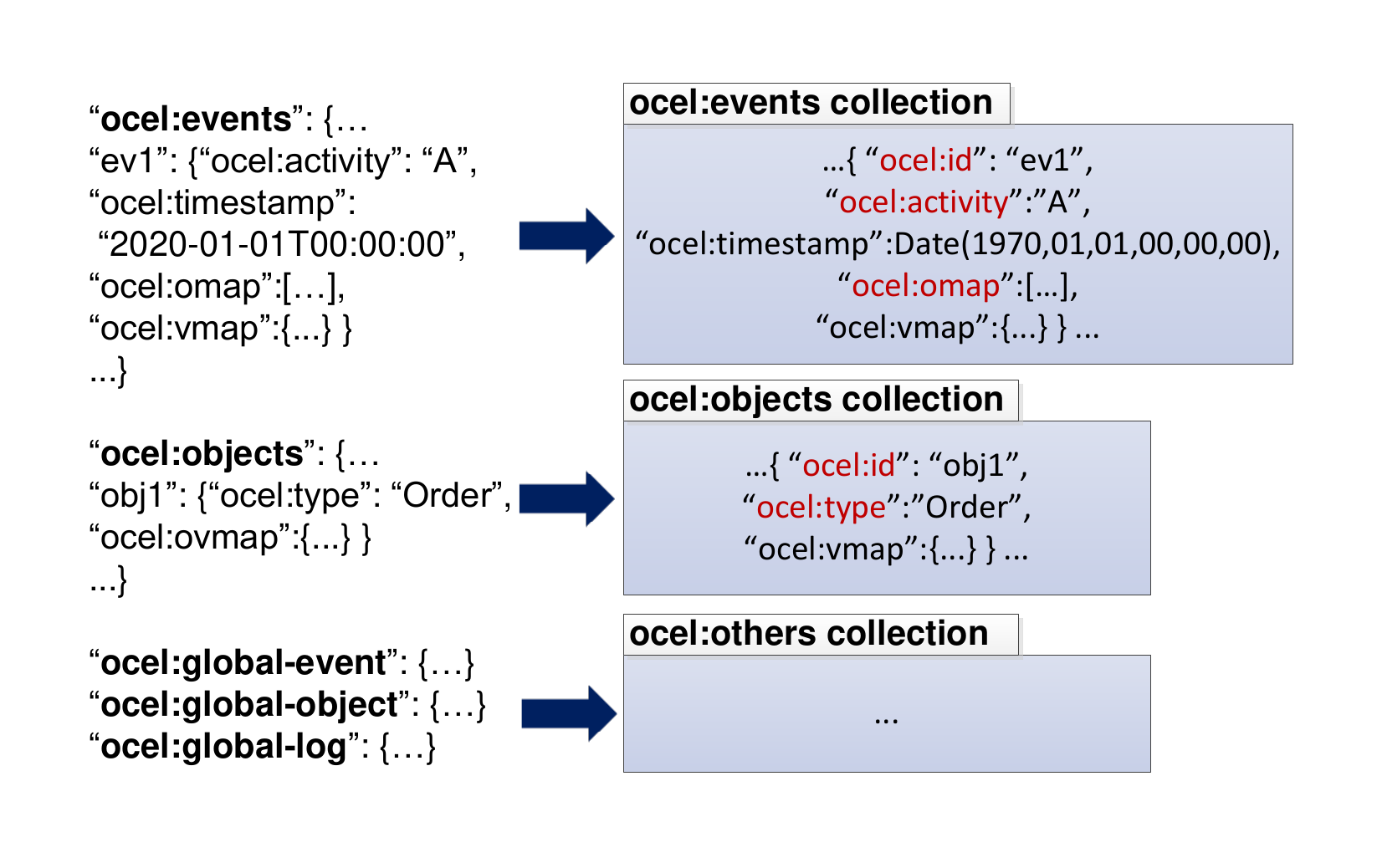}
\vspace{-0.5cm}
\caption{JSON-OCEL implementation (left) and its equivalence to the MongoDB OCEL schema (right).}
\label{fig:translationJsonOcel}
\end{figure}

The implementation of the schema to host the elements of the OCEL standard follows from the implementation of JSON-OCEL \url{http://www.ocel-standard.org/1.0/specification.pdf}. Fig. \ref{fig:translationJsonOcel} shows how the translation of the different entities is possible.


Some fields are colored in red, meaning that an index has been applied to the fields to optimize the execution of some queries. In particular, the identifier, the activity, and the object map (multikey index) have been set as an index for the events. In contrast, the identifier and the type have been set as identifiers for the objects.
The tool permits ingestion of logs in the JSON/XML-OCEL formats or exporting of the MongoDB implementation's contents to JSON/XML-OCEL. Moreover, some essential object-centric process mining operations
have been implemented at the MongoDB level (retrieving the lifecycle of the objects, providing statistics on the number of events, unique and total objects, counting the events per activity and the objects per type \ldots) to reduce the data exchange with the database and use the aggregation features of MongoDB. These are illustrated later in this extended abstract.

\begin{table*}[!t]
\caption{Assessment of the scalability of MongoDB as OCEL storage, for logs of different size.
}
\centering
\begin{tabular}{|l|cccc|cccc|}
\hline
~ & \multicolumn{4}{|>{}c|}{{\bf Insertion + Indexing}} & \multicolumn{4}{|>{}c|}{{\bf mDFG calculation}} \\
\hline
{\bf Size} & Time(s) & JSON(MB) & BSON(MB) & Index(MB) & Ex.time(s) & CPU(\%) & RAM(MB) & Disk usage(MB) \\
\hline
1 M & 79.64 & 610 & 210 & 182 & 25.50 & 64.9 & 883 & 109 \\
5 M & 353.61 & 2805 & 1094 & 875 & 117.65 & 99.9 & 3414 & 501 \\
20 M & 1520.34 & 10899 & 4244 & 3367 & 590.30 & 99.9 & 6705 & 6499 \\
100 M & 7983.87 & 54642 & 21304 & 16852 & 3657.00 & 99.9 & 6709 & 36478 \\
\hline
\end{tabular}
\label{tab:scalabilityAssessment}
\vspace{-3mm}
\end{table*}

\section{Usage of the Tool}

The provided tool is based on the Python language and supports all existing OCEL implementations (JSON, XML, and MongoDB). The tool is available at the address \url{https://github.com/OCEL-standard/ocel-support}.
In particular, example scripts for the usage of the MongoDB interface are available in the folder \emph{examples/mongodb}. First, the connection string and the database name could be set in the script
\emph{commons.py}. The script \emph{exporting.py} permits to load an existing JSON/XML-OCEL file in the MongoDB database, while the script \emph{importing.py} permits to save the object-centric event log
to a JSON/XML-OCEL file. Other scripts perform computations on object-centric event logs:
\begin{itemize}
\item \emph{obj\_centr\_dfg.py} provides routine for the computations of the directly-follows graph for each object type of the log.
\item \emph{activities\_stats.py} and \emph{ot\_stats.py} provide some basic statistics for the activities (number of events and objects) and the object types (number of objects per type) of the event log.
\item \emph{times\_between\_activities.py} provides some statistics of the time passed between a couple of the activities of the log (regardless of the object type).
\end{itemize}

MongoDB offers a powerful aggregation package that permits performing significant object-centric process mining operations directly at the database level.
As an example of a crucial object-centric process mining operation, we show an aggregation that is useful for the computation of the multi-directly follows graph
(finding the events that belong to the lifecycle of an object).
First, the \emph{ocel:omap} attribute (list of related objects) is unrolled, so the same event is
replicated for all the related objects. Then, a grouping operation based on the unrolled \emph{ocel:omap} attribute is performed to collect the activities of the events related to the same object.

\lstset{basicstyle=\small}
\begin{lstlisting}[language=Python]
events_collection.aggregate(
[{"$unwind": "$ocel:omap"},
{"$group": {'_id': '$ocel:omap',
'lifecycle': {"$push": '$ocel:activity'}}}],
allowDiskUse=True)
\end{lstlisting}

The output of the aggregation can be used to calculate the directly-follows graph for the objects of a specific type, and looks like:

\lstset{basicstyle=\footnotesize}
\begin{lstlisting}
[{"_id": "o1", "lifecycle": ["Create Order",
"Payment"]},
 {"_id": "o2", "lifecycle": ["Create Order",
 "Change Order", "Cancel Order"]},
 {"_id": "i1", "lifecycle": ["Emit Invoice",
 "Record Payment"]} ...
]
\end{lstlisting}

\section{Maturity of the Tool}

The prototypal tool available at the address \url{https://github.com/OCEL-standard/ocel-support} has not been used in any real-life case study.
We analyzed the scalability of the MongoDB implementation.
All the experiments have been conducted with a notebook having an I7-7500U CPU, 16 GB of RAM, and an SSD hard drive. Table \ref{tab:scalabilityAssessment} reports on the results attained from logs of different size.
The binary compression used to store the documents by MongoDB permits to save a significant amount of disk space in the storage of the log. We can also see that the index, which is necessary to increase the speed of the computations, occupies a significant amount of space compared to the size of the collection. In the computation of mDFGs, we can see that MongoDB mixes in-memory calculations with on-disk ones, especially if the amount of memory needed is higher than the amount of memory available.
Compared to an in-memory approach, where the entire JSON object is imported into the memory, the computation of the object-centric directly-follows graph takes significantly more time. However, the amount of memory required to store the JSON is also considerably higher than the memory requirements of MongoDB. Our workstation went out of memory trying to ingest an event log having 6.8 M events, while MongoDB can manage bigger logs, as our experiments show.
A video displaying the ingestion of an object-centric event log in MongoDB, and the execution of some computations, is available at the address \url{https://www.youtube.com/watch?v=vDd5CASy1Y0}.

\vspace{-0.2cm}
\section{Acknowledgments}
\vspace{-0.2cm}

We thank the Alexander von Humboldt (AvH) Stiftung for supporting our research. Funded by the Deutsche Forschungsgemeinschaft (DFG, German Research Foundation) under Germany's Excellence Strategy–EXC-2023 Internet of Production – 390621612.

\vspace{-0.5cm}

\bibliographystyle{IEEEtran}
\bibliography{mongodb_ocel}

\end{document}